# Mysterious High Energy Gamma Rays Might Help Explain What Drives Solar Cycles

Gregory S Glenn     January, 2019

**ABSTRACT**

This paper is in response to a technical paper, entitled "Evidence for a New Component of High-Energy Solar Gamma-Ray Production" (Linden, et al., 2018). An article in Scientific American entitled **"**The Sun Is Spitting Out Strange Patterns of Gamma Rays—and No One Knows Why" is a discussion of Linden's paper. It may be summarized as follows: The Sun has been observed to be emitting gamma ray bursts. The weaker gamma rays tend to be less than 50 GeV, emitted during the most active energetic period of the solar cycle and towards the poles. The gamma-ray emission is most intense during Solar Minimum, reaching >100 GeV and those emissions are near the equator: "Most strikingly, although 6 gamma rays above 100 GeV are observed during the 1.4 years of solar minimum, none are observed during the next 7.8 years (Linden, et al., 2018*).*" Pease and Glenn, in the conclusion of a recent paper, suggested that solar cycles are regulated by planetary orbital positions, influencing the Sun through transfer of gravitational or electromagnetic forces, or both (Pease & Glenn, 2016). This paper will describe a working hypothesis that points strongly to electromagnetic connections between Jupiter, Saturn, and the Sun during Solar Minimum which contribute to the high gamma-ray energy observed being emitted by the Sun. The hypothesis further suggests that the electromagnetic connections between the Sun and Jupiter, Saturn, and other planets with magnetospheres, namely Neptune, Earth, and Uranus, are responsible over billions of years for modulating a dual electromagnetic field resonance internal to the Sun. These major periodic cycles are known as the 11-year Schwabe and 22-year Hale solar cycles.

## 1. INTRODUCTION

It is not established why the 11-year Schwabe or 22-year Hale Cycles occur when they do. The predominant theory is that solar cycles are modulated by the Sun's magnetic fields, which are themselves generated by what is called a dynamo effect caused by the sun's rotation. The dynamo effect is due to the Sun, being made of plasma, having a differential spin and not rotating as a solid. It rotates faster at the equator than at the poles and thus the magnetic plasma gets dragged around at different rates. Existing theories do not fully explain why there is a fairly constant internal cycle of approximately 22 years, not based on outside influences.

There are, however, alternative hypotheses about the timing of solar cycles. For example, some in the astrophysics community believe that the movement of sunspots from the Sun's equator to the poles and back is influenced by the orbits of the planets, primarily the massive gas planets Jupiter and Saturn, with some additional influence from the less massive planets.

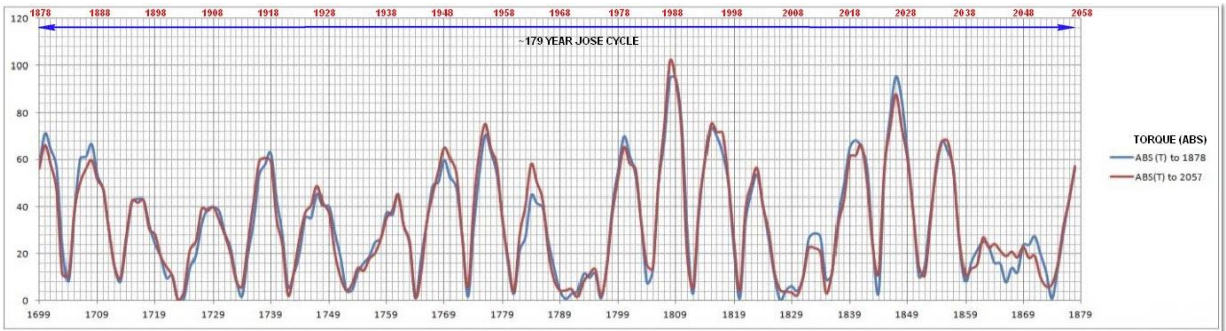

**Figure 1. Torque is very coherent and repeatable in 2 Jose Cycle periods of 179 years each** (G. Pease, et al., 2016)

The transfer of angular momentum to the Sun and how it affects solar activity has been explained both by spin-orbit coupling (I. Wilson, et al., 2008) and by toroidal to poloidal field helicity oscillations (F. Stefani, et al., 2016). Stefani demonstrates how very weak forces can lead to resonance after enough time has passed.

Although Torque from gravitational force (angular momentum transfer) appears to be coherent over multiple time periods (see Figure 1), gravitational forces are not always coherent with solar sunspot cycles. As seen in Figure 2 comparing Torque with solar activity, there are periods of non-coherence, for example during the Maunder and Dalton Minimums.

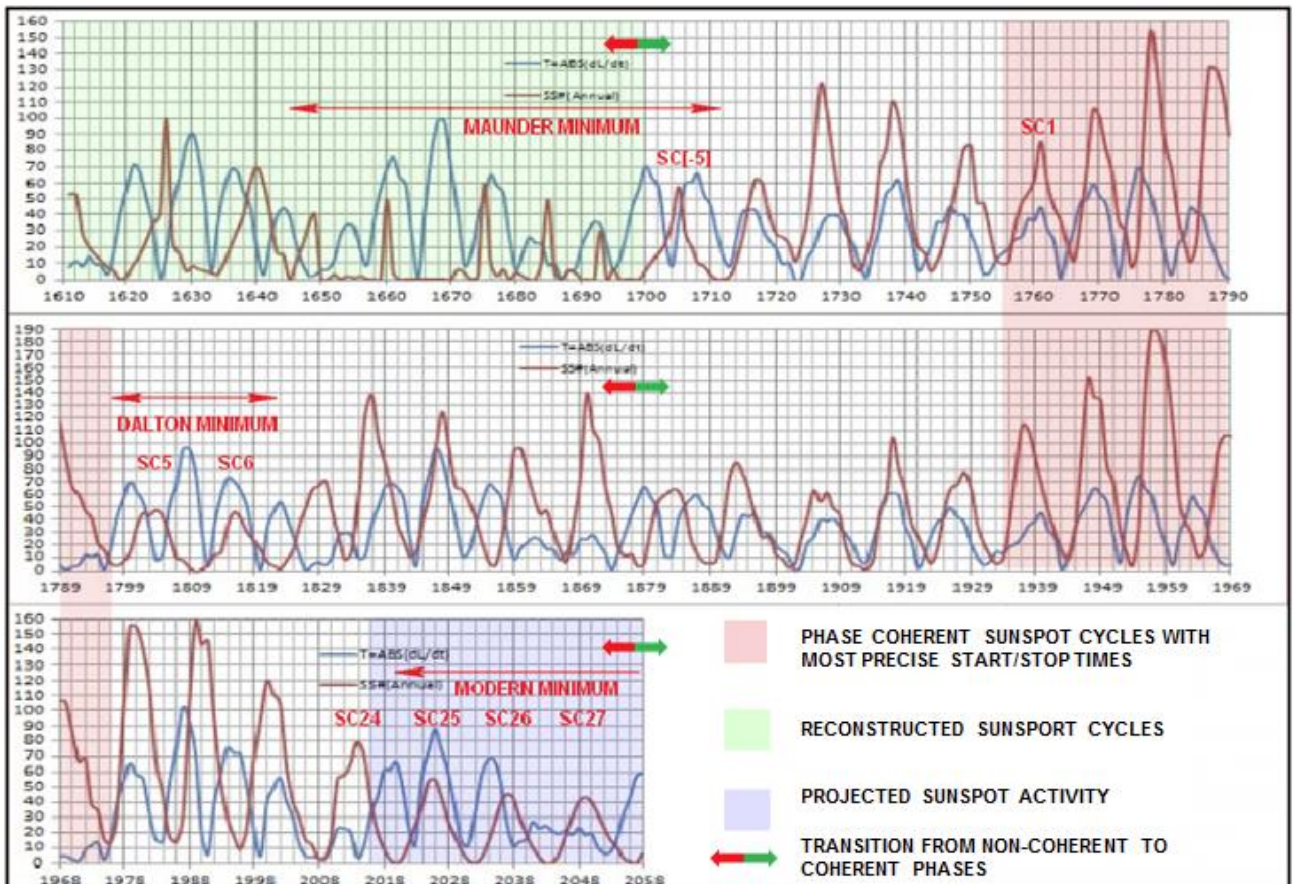

**Figure 2. Torque (blue line) vs Sunspot Number (red line)**         (G. Pease, et al., 2016)
There are periods when Torque and sunspot cycles are coherent and periods when they are not.

The two most massive planets, Jupiter and Saturn, have powerful magnetic fields and have been observed to connect electromagnetically with the Sun through what is known as flux ropes during flux transfer events (FTEs). It is therefore conceivable that electromagnetic forces are influencing the heliomagnetic polar fields and the solar sunspot cycles.

The Fermi Gamma Ray Space telescope has seen powerful 100 GeV gamma ray emissions headed to Earth when there was a solar flare on the back side of the Sun. The research scientists say this is related to coronal mass ejections (CMEs) traveling around the Sun and feeding energy to sunspots on the Earth facing side. CMEs are usually the most powerful during Solar Maximum. Why the strongest gamma ray emissions are during Solar Minimum is a mystery: The cause of this pole/equator shift in gamma-ray emission remains unknown. "I tried to find an interpretation and I came up—one of the few times in my life—totally without any explanation," Jokipii says (S. Hall, 2018).

The observations thus far are that the strongest gamma emissions are found at Solar Minimum, when solar activity (sunspot count and coronal mass ejection quantity and magnitude) is at a minimum. We hypothesize that electromagnetic connections between Jupiter, Saturn, Earth and the Sun influence these gamma ray emissions during Solar Minimum. Perhaps it can be shown that flux transfer events (aka "flux ropes") from Jupiter and Saturn link up and are most intense during Solar Minimum, when high velocity solar wind streams from solar equatorial coronal holes.

## 2. INVESTIGATION

### 2.1 Magnetic Flux Ropes and Birkeland Currents

We first investigate whether or not planet-Sun electrical connections exist. Recent measurements by various spacecraft (Cassini, THEMIS, ESA Cluster, MMS, etc.) have detected flux transfer events of huge amounts of current and voltage flowing between the Sun and some of the planets in the solar system. These so-called "Solar Magnetic Flux Ropes" (B. Filippov, et al., 12 Jan., 2015) emanating from the Sun, are linked back to the Sun by a combination of electric current and magnetic fields moving along solar magnetic field lines, also known as the Parker Spiral. These magnetic ropes consist of counter-streaming flows of energetic particles. They can flow over long distances, as great as those between the Sun and planets. Magnetic flux ropes have certain similarities with planetary Birkeland currents, which are generated by solar wind current flowing from a planet's magnetosphere and connecting with its ionosphere. More than a million amperes have been measured at Earth and many times that at Jupiter. Birkeland currents have some interesting properties allowing them to flow coherently. Per Wikipedia:

"Birkeland currents are also one of a class of plasma phenomena called a z-pinch, so named because the azimuthal magnetic fields produced by the current pinches the current into a filamentary cable. This can also twist, producing a helical pinch that spirals like a twisted or braided rope… Pairs of parallel Birkeland currents will also interact due to Amperes force law: parallel Birkeland currents moving in the same direction will attract each other with an electromagnetic force inversely proportional to their distance apart… Electrons moving along a Birkeland current may be accelerated by a plasma double layer. If the resulting electrons approach the speed of light, they may subsequently produce a Bennett Pinch, which in a magnetic field causes the electrons to spiral and emit synchrotron radiation that may include radio, visible light, x-rays and gamma rays.
(https://en.wikipedia.org/wiki/Birkeland_current)

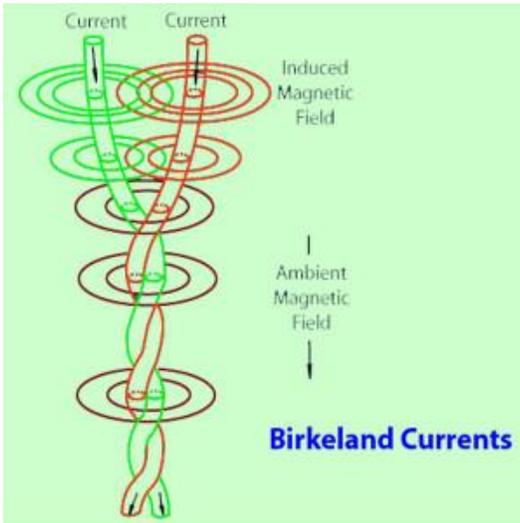

**Figure 3.  Depiction of Birkeland Current**
Helical pinch spiral discourages dispersion and allows current propagation over long distances

The formation of twisted currents helps explain how magnetic flux ropes can travel between the Sun and far reaching planets in the solar system.  Flux ropes have been observed by satellites at several planets and they are observed to connect with many planets similarly to how a CME connects with Earth.

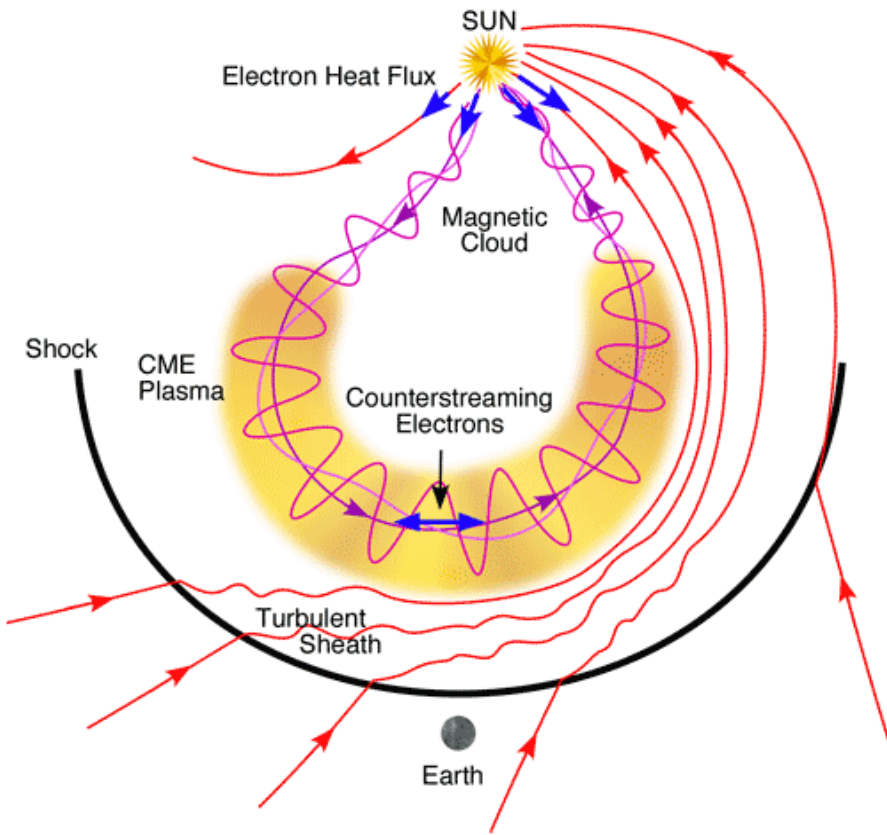

(Zurbuchen & Richardson 2006)

**Figure 4.   Depiction of CME Illustrates how the flow of counter-streaming electrons move between the Sun and a planet's magnetosphere**

## 2.1.1 Magnetic Flux Ropes Connecting Saturn and the Sun

Spacecraft observations confirm flux ropes connecting the Sun and those planets with strong magnetic fields. For example, a Saturn electrical connection to the Sun has been observed by the Cassini spacecraft:

"A twisted magnetic field structure, previously never seen before at Saturn, has now been detected for the first time, using instrumentation built at UCL and Imperial College… These twisted helically structured magnetic fields are called flux ropes or 'flux transfer events' (FTEs) and are observed at Earth and even more commonly at Mercury… This not only shows that magnetic reconnection occurs at Saturn but also that Saturn's magnetic field can at times interact with the Sun in much the same way as at Earth." (UCL, 2016)

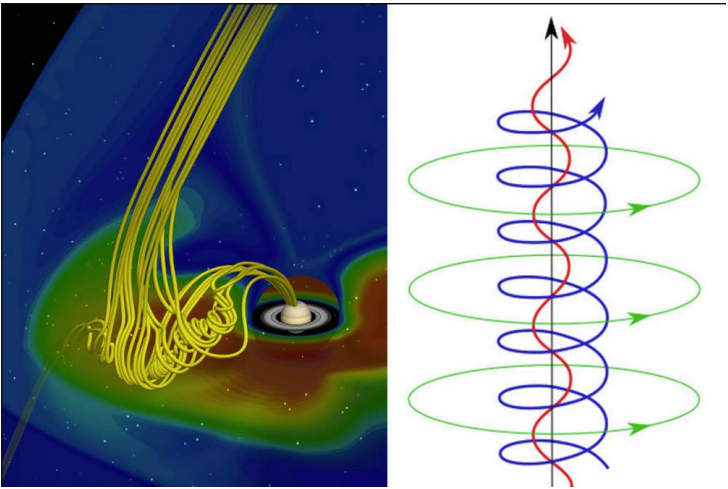

**Figure 5. Magnetic flux rope connecting Saturn and the Sun** (UCL, 6 July, 2016)

## 2.1.2 Magnetic Flux Ropes Connecting Jupiter and the Sun

Research into early magnetic field observations at Jupiter by Pioneer 10 and 11 and Voyager 1 and 2 found evidence of 14 flux transfer events (R. Walker, C. Russell, August 1985).

Magnetic ropes are connected back to the source through the interplanetary magnetic field; thus a reconnection event, essentially a short circuit, sends energy back to the Sun. The energy involved is extremely large. Millions of Amperes have been measured during similar events connecting Earth to the Sun and between Jupiter and Io.

"An electrical interaction between Jupiter and its moons means that they are charged bodies and are not electrically neutral. Jupiter exists in a dynamic electrical relationship to the Sun. As solar max begins, Jupiter's aurorae are becoming more active, with bursts of extreme ultraviolet light detected. The light bursts from Jupiter's aurorae are said to be the result of 'magnetic reconnection' events. The solar wind is said to 'stretch' its magnetic field like a rubber band. When it 'snaps back' the over-stretched magnetic field lines explode, converting some of the 'magnetic energy' to heat and light" (S. Smith, 2014).

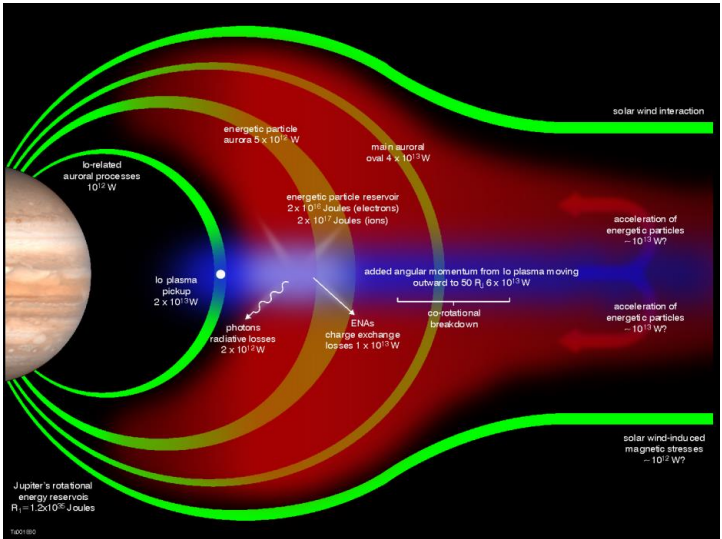
**Figure 6. Solar Wind Interaction with Jupiter** (R. Gladstone, "The Sun-Jupiter Connection")

### 2.1.3 Magnetic Flux Ropes Connecting Earth and the Sun

Earth's electrical connections to the Sun have been observed:
"Researchers have long known that the Earth and sun must be connected. Earth's magnetosphere (the magnetic bubble that surrounds our planet) is filled with particles from the Sun that arrive via the solar wind and penetrate the planet's magnetic defenses. They enter by following magnetic field lines that can be traced from terra firma all the way back to the sun's atmosphere … how FTEs form: On the dayside of Earth (the side closest to the sun), Earth's magnetic field presses against the suns magnetic field. Approximately every eight minutes, the two fields briefly merge or "reconnect," forming a portal through which particles can flow. The portal takes the form of a magnetic cylinder about as wide as Earth."
https://science.nasa.gov/science-news/science-at-nasa/2008/30oct_ftes/

The ends of the flux tube are attached to the solar source and both the magnetic field and electric currents are providing bi-directional connection between the Sun and the planetary magnetospheres.

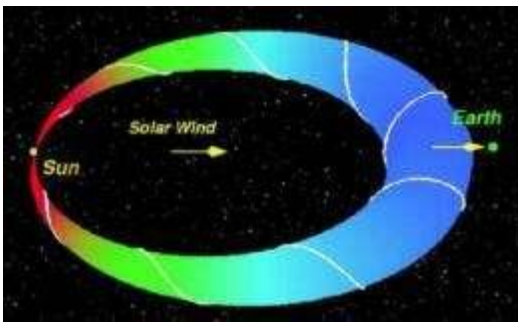 (J. Chen, V. Kunkel, A.P.S.)
**Figure 7. Artist's rendition of an expanding model CME flux rope, about to impinge on the Earth**

### 3. DISCUSSION

### 3.1 How Gamma Rays Could Form at the Sun's Equator

It has been confirmed through satellite instrumentation that Jupiter, Saturn, and Earth have powerful electrical connections with the Sun. Our hypothesis needs to provide a mechanism for how the relationship between these magnetically enhanced planets electrically connects with the Sun as to produce energetic >100GeV gamma ray emissions.

During Solar Minimum, sunspots move towards the equator through migration of the polar magnetic fields. The equator is where the Parker Spiral streams out energy towards the planets during Solar Minimum (see Figure 8a).  It now needs to be shown that planetary magnetic connections at the Sun have sufficient energy to produce gamma rays.

One mechanism for producing gamma rays is through the so-called double layer, a structure in a plasma consisting of two parallel layers of opposite electrical charge.  The Alfven-Carlquist double-layer theory of solar flares uses the Vlasov equations for ions and electrons to show that a double layer can form from both free and trapped electrons and ions.  These particles will be accelerated in a double layer and form extremely high-energy particles as seen during solar flares (S. Hasan, D. Ter Haar, December, 1977).  https://link.springer.com/article/10.1007%2FBF00643464

Planetary connections to the Sun may contribute to a double layer and as mentioned earlier, a mechanism exists to produce gamma rays:  "Electrons moving along a Birkeland current may also be accelerated by a plasma double layer.  If the resulting electrons approach relativistic velocities (i.e., if they approach the speed of light) they may subsequently produce a Bennett pinch, which in a magnetic field causes the electrons to spiral and emit synchrotron radiation that may include radio, optical (i.e. visible light), x-rays, **and gamma rays**." (https://en.wikipedia.org/wiki/Birkeland_current)

In fact, synchrotron radiation has been seen around Jupiter, which has a magnetic field almost 20,000 times stronger than Earth's:
"Trapped relativistic electrons (at energies of 10s of MeV) form Jupiter's 'Van Allen belts' within a few radii of the planet.  These electrons emit synchrotron radiation at decimetric (10s of cm) radio wavelengths as they spiral around and bounce from north to south along magnetic field lines.  Jupiter is the only planet that emits synchrotron radiation, although it is often seen elsewhere in the universe."
(http://spacewx.com/jupiter/docs/JupiterAurora.pdf)

The aforementioned Z-pinch (aka Zeta-pinch, a specific type of Bennett pinch) in counter-rotating electron flow (i.e. Birkeland current) attracts matter, which could then be accelerated up to gamma ray intensity at the connection points at the Sun and planet.

**3.2  Evidence of Planetary Electromagnetic Connections Influencing the Solar Cycle**

Planets in our Solar System having the strongest magnetic fields measured at the planet are Jupiter, at 19,519 times stronger than Earth's field, and Saturn, at 578 times stronger than Earth.  The fact that Jupiter and Saturn display strong auroras near their poles indicates that there are strong electromagnetic connections to the Sun, similar to that with Earth.  We have shown that to be the case.

M.A. Vukcevic made an interesting discovery in 2012.  He looked at the separation angle between Jupiter and Saturn at Solar Cycle Minimum in two distinct ways. The first way was the simple geometrical angle made in space (see Figure 9a). The second was what he called 'the magnetospheric angle' (or helio-magnetic angle), which he measured along the Parker Spiral, the spiral of varying strength of the solar wind relative to the stellar background (see Figure 9b).

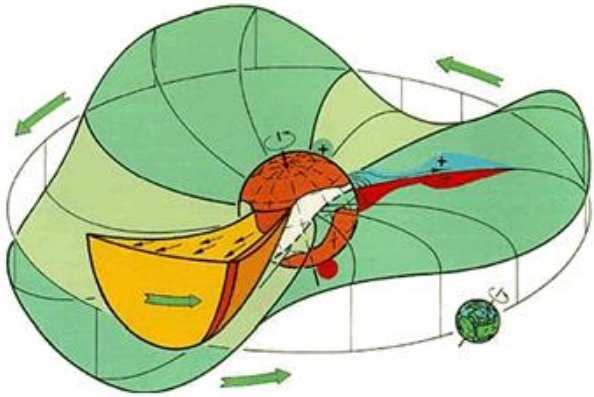 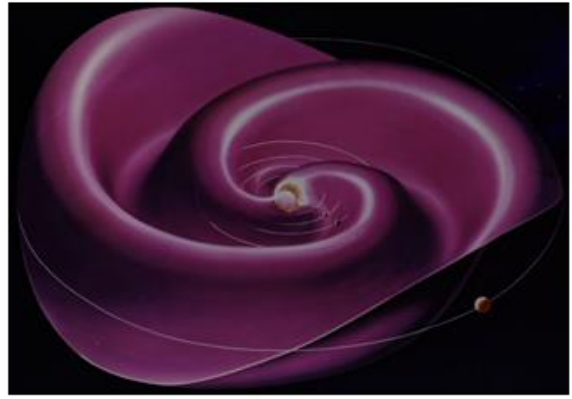

**Figure 8a. Parker Spiral at Solar Minimum**. Note positive (blue) and negative (red) current flow from the Sun's Equator. The Parker Spiral is named after Eugene Parker, its discoverer. (MalagaBay, 2015)
**Figure 8b.** The spiraling magnetic sheets change polarity and warp into a wavy spiral shape similar to a spinning ballerina skirt. NASA (Public Domain)

The Parker Spiral is so shaped because it moves outwards as the Sun spins on its axis. When Vukcevic measured the J-S angle to the Sun along the Parker Spiral at Solar Minimum, it revealed a distinct pattern pointing to a Jupiter-Saturn influence on solar activity.

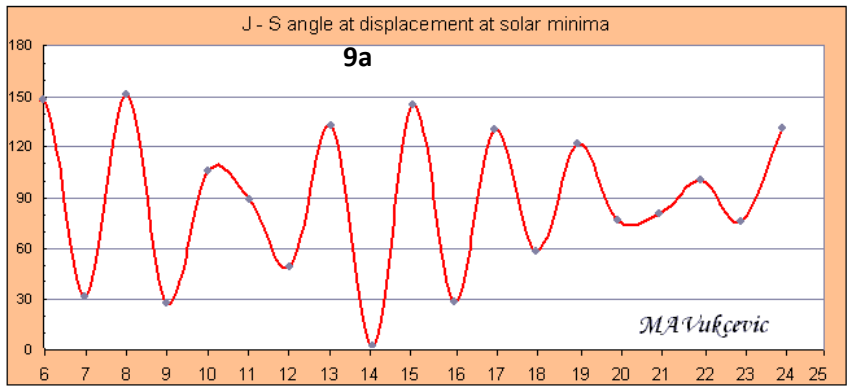
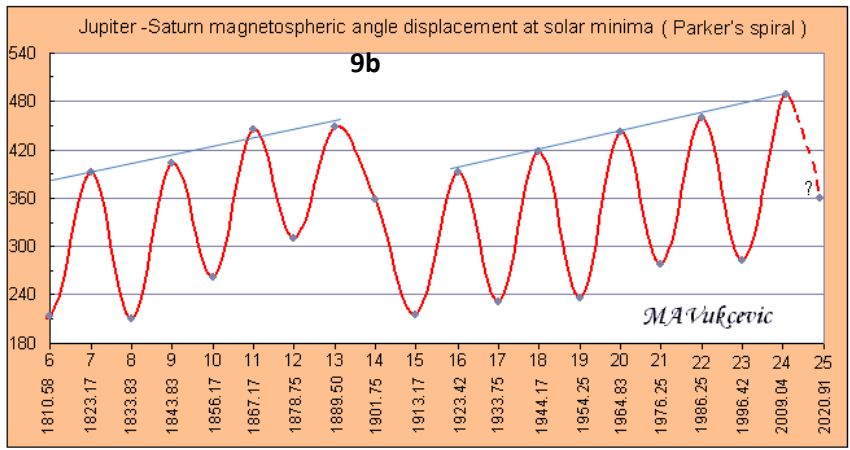

(M.A. Vukcevic, 2012)
**Figure 9a & 9b.** Jupiter-Saturn Helio-magnetic Angle at Solar Minimum (Solar Cycles 6 – 24, SC25 estimated)

The second plot, figure 9b, calculated by measuring angles along the spiral of the electromagnetic Parker Spiral, shows a coherent repeating pattern. This suggests a relationship between the relative positions of Jupiter and Saturn and the occurrence of Solar Cycle Minimum, which is known to be driven by electromagnetic fields.

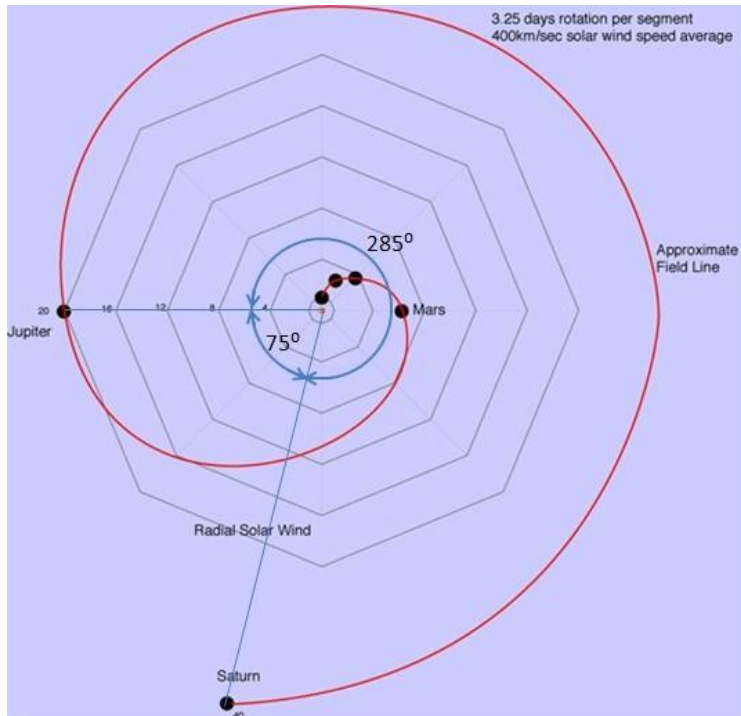

**Figure 10 J-S Angle calculated two ways (annotated)** ( https://www.jupitersdance.com/thefinalwaltz/ )

Figure 10 illustrates how the angle between Jupiter and Saturn can be calculated two ways. The 75° angle is the standard measurement between the planets along the ecliptic, as shown in figure 9a for Solar Cycles 6-24. The 285° angle is measured along the Parker Spiral, as shown in figure 9b. If, for example, Saturn continued past Jupiter into the 12 o'clock position, the standard angle would be 90° in figure 9a and 360+90=450° in figure 9b along the Parker Spiral.

The Vukcevic study prompted the following response: "...All in all, I'm inclined to the view that the timing of solar cycles is being modulated more strongly by electromagnetic forces than by tides or gravity" (R. Tattersall, March 22, 2012).

**3.3 Numerical Discussion of Solar Cycle Length**

H. Schwentek and W. Elling of the Max Planck Institute studied the relationship between statistical spectral bands in sunspot number and the space-time organization of the solar system. Their conclusion was that the dominant sunspot number spectral band, the 10.8 year Schwabe solar cycle, was derived by the configuration period of Jupiter and Saturn (the J-S synodic period of 19.859 years) times the ratio of their distances from the Sun (0.545 years). (H. Schwentek & W. Elling, July, 1984)

The derivation is as follows, as summarized by R. Tattersall:
   T1*T2/(T2+T1) = 23.72 (twice Jupiter's orbital period) times 19.85 (the J-S synodic period) all divided by 23.72 plus 19.85 = **10.806**.

Kepler's third law states: The square of the orbital period of a planet is directly proportional to the cube of the semi-major axis of its orbit.

The orbital ratio can be derived as follows. For the orbits of Jupiter (11.86 years) and Saturn (29.46 years) we find that the squares (multiplication by itself) of the orbital periods are 140.67 and 867.3. The cube roots of these values are 5.2 and 9.54. The ratio of these values (one divided by the other) is 0.545.

As Jupiter passes Saturn at conjunction it then takes just under 20 years for Jupiter to catch up with Saturn again.

We can calculate this using a law discovered by Kepler's mentor Copernicus: The Synodic period is given by the inverse of the inverse of the orbital period of the slower moving body minus the inverse of the orbital period of the faster moving body.

Synodic period of Jupiter and Saturn is $1/(1/11.86 - 1/29.46)$ = **19.852 years.**

We can then multiply that result by the **orbital distance ratio of 0.545** we previously calculated to obtain **10.819.**

The result of 10.819 years is the same time as the Schwabe Cycle. The Hale Cycle is 2x the Schwabe Cycle. This further illustrates an orbital relationship between the electromagnetically active planets Jupiter and Saturn, as seen in figures 11a and 11b.

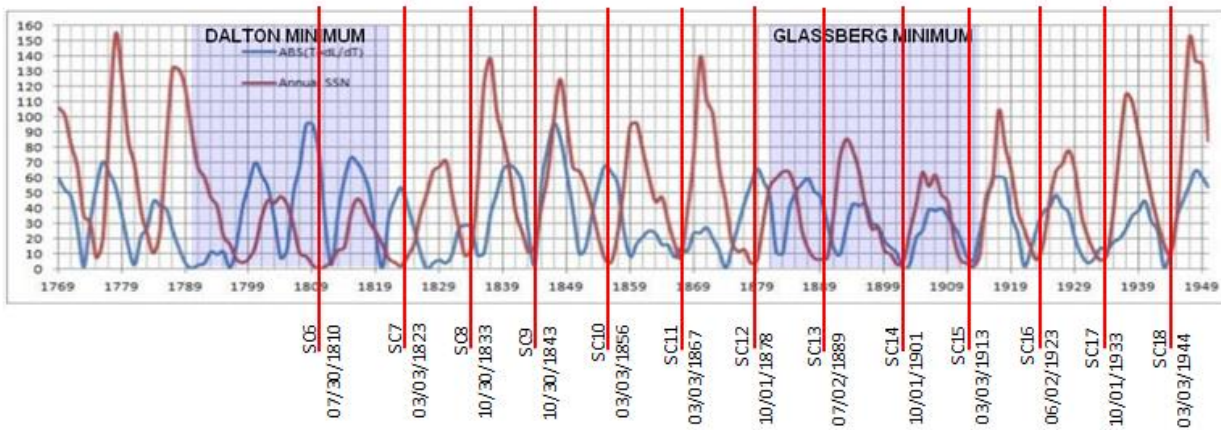

**Figure 11a. Observed relationship between the relative position of Jupiter and Saturn (blue dots in Figure 9b) and Solar Cycle Minimums (SC6 – SC18: red vertical lines and SSN red plot line)** (G. Pease & G. Glenn 2016, annotated)

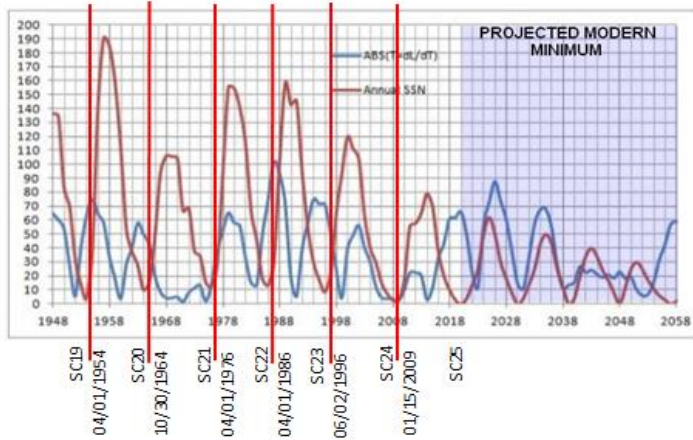

**Figure 11b. Observed relationship between the relative position of Jupiter and Saturn (blue dots in Figure 9b) and Solar Cycle Minimums (SC19 – SC25: red vertical lines and SSN red plot line)**  G. Pease & G. Glenn 2016 (annotated)

### 3.5  Possible Explanation for Higher Energy Gamma Ray Bursts at Solar Minimum

It is known that a particular type of solar activity is highest at Solar Maximum.  There are more sunspots, more coronal mass ejections, and a more energetic solar wind (Parker Spiral); that is why the occurrence of the highest energy gamma rays at Solar Minimum is considered a mystery.  However, there are certain characteristics of Solar Minimum that should not be overlooked.

Large coronal holes appear during Solar Minimum, allowing streams of solar particles to escape at very high velocity, as the magnetic field structures are open and plasma is not trapped by the magnetic fields.  These powerful high velocity particle streams connect to a planet's magnetosphere and can be observed to cause significant geomagnetic storms.  Coronal holes are often seen to migrate from the poles to near the equator during Solar Minimum.  Significantly, Solar Maximum occurs just after the polar fields reverse and polar field strength is near zero.  However, Solar Minimum occurs when the polar field strength is near maximum strength, as can be seen in Figure 12 (red arrows).  At Solar Minimum, sunspots have migrated to near the equator, where energetic gamma rays are observed and when polar field strength is at a maximum.

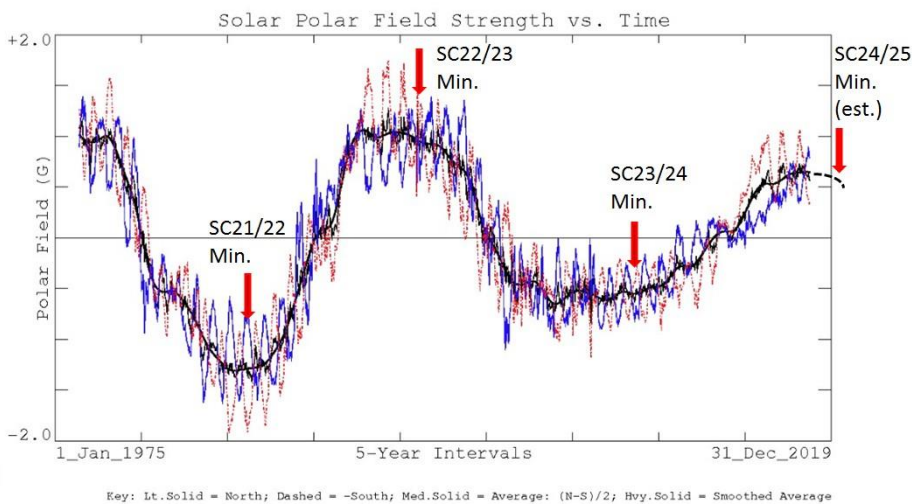

 Wilcox Solar Observatory (annotated)
**Figure12.  Polar fields are near maximum strength at Solar Minimum (red arrows)**

## 4. CONCLUSION

We have provided the beginning of a fairly coherent hypothesis. First, we show that powerful electromagnetic forces flow between Jupiter, Saturn, Earth and the Sun at certain orbital positions. We further show that these connections take on the characteristics of helical spirals similar to Birkeland currents and propose that such currents are capable of generating gamma rays through a mechanism known as a Bennett pinch. Gamma rays have been observed generated naturally at just two locations within the solar system: at the Sun and at Jupiter. Finally, we have shown that a coherent pattern of Jupiter-Saturn-Sun alignment along the Parker Spiral at approximate 10.8 year intervals is related to Solar Minimum. This timing is when gamma rays are observed as most energetic.

So far, these observations are just coincidences and smoking guns. What is needed to prove this hypothesis is an observation of the gun firing: spacecraft measurements in real time showing these powerful electromagnetic connections between the Sun, Jupiter, and Saturn during the next Solar Minimum gamma ray emission.

Since the next Solar Minimum is due soon, the proof might be close at hand. Solar physicists at the Wilcox Solar Observatory have recently predicted the Solar Minimum between Solar Cycles 24 and 25 in a few months from the time of this writing in early 2019. Solar Cycle 25 sunspots have already started to occur. However, according to the hypothesis presented here, the alignment for Solar Minimum SC24/SC25 likely won't occur until much later, likely 2020. Therefore, a judgement for this hypothesis may soon be tested.